\documentstyle[aps,prb,multicol,epsf,graphicx]{revtex}

\newcommand \be {\begin{equation}}
\newcommand \ee {\end{equation}}
\newcommand \eps {\epsilon}
\newcommand \bi {\bibitem}
\newcommand \s {\sigma}
\newcommand \lan {\langle}
\newcommand \ran {\rangle}

\begin{document}

\title{Scaling approach to order-parameter fluctuations in disordered frustrated systems}
 \author{Felix Ritort and Marta Sales}
\address{Departament de F\'{\i}sica Fonamental, Facultat de F\'{\i}sica, Universitat de 
Barcelona\\
Diagonal 647, 08028 Barcelona (Spain)\\
E-Mail: ritort@ffn.ub.es, msales@ffn.ub.es}


\maketitle
\begin{abstract}
We present a constructive approach to obtain information about the
compactness and shape of large-scale lowest excitations in disordered
systems by studying order-parameter fluctuations (OPF) at low
temperatures. We show that the parameter $G$ which measures OPF is
$1/3$ at $T=0$ provided the ground state is unique and the probability
distribution for the lowest excitations is gapless and with finite
weight at zero-excitation energy. We then apply zero-temperature
scaling to describe the energy and volume spectra of the lowest
large-scale excitations which scale with the system size and have a weight at zero energy $\hat{P}_v(0)\sim l^{-\theta'}$ with $v=l^d$.
A low-temperature
expansion reveals that, OPF
vanish like $L^{-\theta}$, if $\theta> 0$ and remain finite for space
filling lowest excitations with $\theta=0$.  The method can be
extended to extract information about the shape and fractal surface of
the large-scale lowest excitations.
\end{abstract}
\begin{multicols}{2}
\narrowtext

{\it Introduction.} There is a long standing controversy on the nature
of low-temperature excitations in spin glasses. This discussion has
seen a strong revival in the last years due to the large increase of
computational capabilities together with the simultaneous development
of more refined and powerful numerical techniques
\cite{PY,KPYHG}. These have largely improved the ability to find exact
ground states or to equilibrate small systems at very low
temperatures. The aim of this letter is to present some analytical
results on this subject by investigating order parameter fluctuations
(hereafter denoted by OPF) and showing how they can give us
information about the topology of the excitations in spin glasses.

The study of parameters measuring OPF has been proven to be very
powerful to locate finite temperature phase transitions in disordered
systems ~\cite{MNPPRZ,ALTRI}. Recent studies
~\cite{PRS} show that these parameters can bring also information
about the frozen phase and thus shed some light on the nature of the
spin-glass phase. The introduction of $G$ was originally motivated by
the existence of the well-known Guerra sum rules for
spin-glasses\cite{GUERRA1}, and in particular the following one:
$\overline{\langle q^2\rangle^2}=\frac{1}{3} \overline{\langle
q^4\rangle}+\frac{2}{3}\overline{\langle q^2\rangle}^2~~~$ where
$q=\frac{1}{V}\sum_{i=1}^V\sigma_i\tau_i$ is the usual overlap between
different replicas $\lbrace \sigma,\tau\rbrace$, $\lan ...\ran$ and
$\overline{(...)}$ stand for thermal and disorder averages
respectively. This relation implies that in systems with non vanishing
OPF in the frozen phase, the parameters $G$ and $A$ defined as  \be
G=\frac{\overline{\langle q^2\rangle^2}-\overline{\langle
q^2\rangle}^2} {\overline{\langle q^4\rangle}-\overline{\langle
q^2\rangle}^2}~~,A=\frac{\overline{\langle q^2\rangle^2}-\overline{\langle
q^2\rangle}^2}{\overline{\langle q^2\rangle}^2}
\label{eq2}
\ee have the following behavior in the infinite-volume limit:
$G=\frac{1}{3}\theta(T_c-T)$ and $A\equiv \hat{A}(T)$ is a non-vanishing
function of $T$ (this result for $A$ being an evidence of replica-symmetry
breaking (RSB)).

Very recently it was conjectured in ~\cite{RS}~(henceforth referred
to as RS) that for systems of Ising variables $G$ takes the
universal value 1/3 at $T=0$ for finite volume systems under two
mild assumptions: the uniqueness of the ground state and the absence
of zero gap in the local field distribution. In RS the main scenarios
for OPF were discussed and many examples were given to check the
validity of such a conjecture. 
Moreover, the conjecture was proven for finite volume systems provided
that the limit $T\to 0$ is taken before the limit $V\to\infty$ by showing
that one-spin excitations with finite probability at zero local field
yield at low temperature the dominant (linear in $T$) contribution to
OPF. 

In this letter we prove that these results generally hold whatever the
size of the lowest excitations is, provided these are gapless and have
finite weight at zero excitation energy. Moreover we will obtain
valuable information about the size of the excitations by considering
the low-temperature behavior of the OPF parameters.  In particular, we
will see how we can establish a scaling theory for the lowest
large-scale excitations which relates OPF to the $T=0$ thermal scaling
exponent $\theta'$ and the exponent which characterizes the distribution of large-volume excitations $\gamma$.
We exemplify these findings in the
spin-glass chain. In this paper we analyze the excitations through the overlap parameter $q$ and we do not discuss about the shape of the excitations, an issue which 
be addressed by studying the energy overlap or link overlap
order parameters in a similar way.

{\it The model.} Consider a system of $V$ Ising spins with a general Hamiltonian

\be
{\cal H}=-\sum_{i,j} J_{ij} \s_i \s_j ~~~,
\label{eq4}i
\ee where the $J_{i,j}$ are the quenched variables. In general these
can be either $0$ (depending on the dimensionality and interaction
range of the system) or distributed according to some continuous
$P(J)$. For a continuous distribution of couplings there is a unique
ground state (up to a global flip of all the spins), whose
configuration we denote by $\{\s_i^*\}$. This Hamiltonian can be
rewritten in terms of local fields, following the formulation given in
RS, ${\cal H}=-\sum_{i}\s_i\,h_i~$.  In the ground state configuration
 each spin $\s_i$ is aligned with its local field $h_i$,
so that the ground-state energy is $E_{\rm GS}=-\sum_{i}\s^*_i
h^*_i=-\sum_{i}|h^*_i|=-\sum_{i,j} J_{ij} \s^*_i \s^*_j$.  Suppose
that for a given sample we know the ground state configuration
$\{\s_i^*\}$. In this state, we will take the convention that all the
spins have the same color, let say red. Consider now a cluster
excitation of volume $v$. In this new state $v$ spins will change
their orientation with respect to the ground state configuration so
that they have a different color, let say blue.  Thus, in the
excited state the spins will fall into two groups: the red one, which
we denote by ${\cal D}$, and the blue one, which we denote by
$\overline{{\cal D}}$.  This minimum cost excitation is always a
cluster excitation in the sense that all excited spins are connected
to each other by bonds. An excitation consisting in two or more
disconnected clusters will necessarily cost more energy than reverting
one single cluster. Since we are considering a cluster excitation, we
can define among these two regions the surface of the excitation or
droplet: $\delta {\cal D}$.  $\delta {\cal D}$ will be constituted by
those pairs of spins of different color which are connected to each
other by at least one bond. Thus the surface contains red ($\delta
{\cal D}\cap{\cal D} $) and blue spins ($\delta {\cal
D}\cap\overline{{\cal D}}$). 

The energy cost of such an excitation
will depend exclusively on the bonds which connect spins of
different colors, defining the surface of the droplet. Therefore, the
energy cost reads,

\be
\Delta E= 2\sum_{\stackrel{\rm i,k~diff. color}{(i,k)\eps \delta {\cal D}}} J_{i,k}\s_i^*\s_k^*~~~,
\label{eq6} 
\ee 
where each coupling is counted only once.  It follows that the
lowest excitation with energy
($\Delta E ^*$) will correspond to breaking a subset of bonds
among the $\{J_{i,k}\}$ such that the value in (\ref{eq6}) is minimum.


{\it Analysis of OPF fluctuations.} If  we are at a very
low temperature the main contribution to the
partition function arises from the lowest excitation.  In this limit
the partition function reads

\be
{\cal Z}= e^{\tiny -\beta E_{\rm GS}}~\left(1+e^{\tiny -\beta \Delta E^*}\right)~~~~~.
\label{eq9}
\ee

The computation of the terms involved in $G$ is done following the same
procedure used in RS. Different quantities which enter in
the definition of $G$ can be expressed in terms of the two and four
point correlation functions by defining the following objects
$T_{ij}\equiv\lan\s_i\s_j\ran^2$ and
$T_{ijkl}\equiv\lan\s_i\s_j\s_k\s_l\ran^2$.

\begin{eqnarray}
\langle q^2\rangle &=& \frac{1}{V}+\frac{1}{V^2}\sum_{i\ne
j}T_{ij}~~~~;
\langle
q^4\rangle = \frac{3}{V^2}-\frac{2}{V^3} +
\nonumber\\&&\bigl(\frac{6V-8}{V^4}\bigr)\sum_{i\ne j}T_{ij}
+\frac{1}{V^4}\sum_{(i,j,k,l)}T_{ijkl}~.
\label{eqq4}
\end{eqnarray}


To compute the two-point correlation functions $\lan\s_i\s_j\ran$ 
we must take into account two different contributions which we denote
as $U_{ij}$ and $V_{ij}$:

\begin{itemize}
\item Pairs of spins of the same color, either blue $\s_{i(j)}=-\s_i^*$,
or red $\s_{i(j)}=\s_i^*$, both yield the same contribution $\lan \s_i\s_j\ran= \s_i^*\s_j^*\equiv U_{ij}$.

\item Pairs of spins having different colors yield $\lan \s_i\s_j\ran= \s_i^*\s_j^*\tanh \left({\tiny \frac{\beta \Delta E^*}{2}}\right)\equiv V_{ij}~~.$

\end{itemize}

In the same way the  four-point correlation functions $\lan\s_i\s_j\s_k\s_l\ran$ can be
computed by adding two types of contributions: $U_{ijkl}$ and $V_{ijkl}$.
\begin{itemize}
\item When either, all four spins have the same color or there are two
spin pairs with different colors, $\lan \s_i\s_j\s_k\s_l\ran= \s_i^*\s_j^*\s_k^*\s_l^*\equiv U_{ijkl}~~.$

\item When three spins have the same color and the other spin has a
different color $\lan \s_i\s_j\s_k\s_l\ran= \s_i^*\s_j^*\s_k^*\s_l^*~\tanh \left({\tiny \frac{\beta \Delta E^*}{2}}\right)\equiv V_{ijkl}~~.$

\end{itemize}

The next step in order to compute $\overline{\langle q^2\rangle}$,
$\overline{\langle q^2\rangle^2}$ and $\overline{\langle q^4\rangle}$ is
to take into account the degeneration $d_v$ of each kind of contribution. This
number will depend on the size of the excitation ($v$).  Consider, for
instance, the term $\sum_{i\neq j} T_{ij}$.
An excitation with $v$ blue spins and $(V-v)$ red ones contributes to
$T_{ij}$ with the following degenerations:
\begin{itemize}
\item The number of pairs of spins which in the excited state are both red 
($d_v=(V-v)(V-v-1)$) or both blue ($d_v=v(v-1)$) yield $U_{ij}^2=1$.   
\item The number of pairs of spins with different colors ($d_v=2(V-v)v$)
yield $V_{ij}^2\equiv R\equiv \tanh^2 ({\tiny\frac{\beta \Delta E^*}{2}})$.  
\end{itemize}
Adding all contributions we finally obtain: $\sum_{i\ne j}T_{ij}=(V-v)(V-v-1)+v(v-1)+2(V-v)vR$ .
Up to now we have contented ourselves with computing the correlation
functions but we have not made any hypothesis on the probability
distribution of having an excitation of certain volume ($v$) and
energy cost ($\Delta E^*$).  When performing the average over the
disorder, we have to bear in mind that each sample may have a lowest
excitation of size $v$ and energy $\Delta E^*$, yielding a
distribution of the lowest excitation energies for every size $v$.
Since there may exist correlations among fields which enhance clusters
of certain sizes we make the assumption that excitations of volume
$v~(<V)$ can take place with a finite probability $g_v$
($\sum_vg_v=1$) and the energy cost of these excitations at fixed
volume $v$ is distributed according to some normalized continuous $\hat
P_v(\Delta E^*)$, which is only defined for positive $\Delta
E^*$. Through these considerations we get the following general
expression $\sum_{i\ne j}\overline{T_{ij}}=\sum_{v=1}^{V-1} g_v
\bigl[V(V-1)+2v(v-V)+2 (V-v)v~ \overline{R_v^1}\bigr]$ where
$\overline{R_v^k}= \int d \Delta E^*\hat{P_v}(\Delta
E^*)\left(\tanh({\tiny \frac{\beta \Delta E^*}{2}})\right)^{2k}~~.$


In a similar way we can evaluate all terms entering in (\ref{eqq4}). After some
algebra we obtain for the numerator ($N$) and denominator ($D$) of $G$
the following expressions,
\begin{eqnarray}
N=\frac{4}{V^4}[\sum_v g_v(1-2\overline{R_v^1}+\overline{R_v^2})v^2(v-V)^2\nonumber\\-\bigl(\sum_v g_v(\overline{R_v^1}-1)v(V-v) 
\bigr)^2]~~~~,
\label{eq14}\\
D=\frac{4}{V^4}[\sum_v g_v(1-\overline{R_v^1})2v^2(v-V)^2\nonumber\\-\bigl(\sum_v\,g_v(\overline{R_v^1}-1)v (V-v)\bigr)^2]~~~~.
\label{eq15}
\end{eqnarray}

Note that these expressions are invariant under the change $v\to V-v$
as it should be (for every excitation there is the corresponding one
related by time-reversal symmetry). Now we compute the low-temperature
limit of the quantities $\overline{R_v^k}$ with $k=1,2$. A simple
calculation yields, in the limit $T\to 0$ keeping terms up to first
order in $T$, $\overline{R_v^1}=1- 2\,T \hat P_v(0)+ {\cal O}(T^2)
~~,\overline{R_v^2}=1- \frac{8}{3}T\, \hat P_v(0) +{\cal O}(T^2)~~.$
Substituting them into expressions (\ref{eq14}),~(\ref{eq15}), we
obtain that both numerator and denominator vanish linearly with $T$
but its ratio remains finite yielding the expected result,

\begin{eqnarray}
G=\frac{1}{3}+{\cal
O}(T)~~~,\label{eq17a}\\
A=\frac{16T}{3V^4}\sum_{v=1}^Vg_v\hat{P}_v(0)v^2(V-v)^2+{\cal O}(T^2)~~~~.
\label{eq17b}
\end{eqnarray}

 The only necessary condition to obtain
these last results is that $\hat P_v(0)$ has a finite weight at zero energy cost for
one given volume ($v$). Otherwise, the expression for
$\overline{R_v^k}$ would have corrections of order ${\cal O}(T^2)$ and
the universal value $1/3$ for $G$ and the linear in $T$ dependence of
$A$ would not be recovered anymore~\cite{RS}. Let us note that in RS
we considered one-spin excitations so we assumed that, $\hat
P_v(0)\neq 0$ for $v=1$. In that work we stressed that in order to have
corrections ${\cal O} (T)$ when considering the inversion of two or
more spins there had to be singular correlations between local fields
at different sites. This is the situation we expect to meet in generic
spin-glass systems where large cluster excitations are responsible for
the universal low-temperature value of $G$.

{\it Scaling theory for OPF fluctuations.} Up to now we only considered
the linear terms in a low $T$ expansion. But this first order term
suffices to establish the scaling behavior of $A$.  In this approach we will focus on large-scale excitations which scale with the volume of the system, assuming that they bring the relevant contribution to thermodynamic quantities.
Let's suppose that the weight at zero energy of lowest excitations scales like  $\hat{P}_{v=l^{d}}(0)\sim l^{-\theta'}$,  where  $d$ is the
dimensionality of the system and $\theta'$ is the
thermal exponent for the lowest excitations \cite{FOOTNOTE3}. Then we  suppose the following scaling ansatz for lowest-lying large-scale excitations:
\be
 g_v=L^{-\gamma}\hat{g}\bigl(\frac{v}{L^{d}}\bigr)~~~,
\hat P_{v=l^{d}}(0)\sim l^{-\theta'}~~,\frac{v}{V}\propto{\cal O}(1) ~~.
\label{scaling}
\ee
where $L$ is the lattice size  and $d\leq \gamma$ (since $g_v$ is positive defined and normalizable). This functional form implies that in order to assure the normalization condition $\sum_v g(v)=1$, the weight of finite volume excitations is $1-L^{-\gamma}$.
Note that, in principle, since excitations of any size are possible, the average volume of the excitations  scales like $\overline{v}\sim L^{2d -\gamma}$~~~~\cite{foot}.
Substituting these relations in $A$ in (\ref{eq17b}) and taking the large $L$ limit we get,
\be A=\frac{16}{3}T
L^{-\gamma+d-\theta'}\int_0^{1}dx\hat{g}(x)x^{2-\frac{\theta'}{d}}\Bigl(1+{\cal
O}\bigl ({x}\bigr)\Bigr).
\label{A}
\ee
 Assuming that $\hat{g}(x)$ decays fast
enough for $x\gg 1$ (e.g. exponentially), the general low $T$ scaling
$A\sim TL^{-\theta}$, $\theta\equiv\theta'-d+\gamma$ together with relations (\ref{scaling})
provide a way to estimate both $\theta'$ and $\gamma$
\cite{FOOTNOTE5}. 
  In this 
scenario, since $A$ is bounded $A<1$, the crossover behavior of $A$
is then determined by the term $TL^{-\theta}$ which gives a typical
length scale $l^*\sim T^{1/\theta}$. In usual domain wall scaling
theory \cite{DROPLET} we expect that, for $\theta<0$, there is a
relation between the thermal exponent $\theta$ which accounts for the
scaling of the free-energy of the largest excitations \cite{FOOTNOTE3}, and
the correlation length exponent $\nu=-1/\theta$.
Below the lower critical
dimension $d_l$ we have $\theta\le 0$. Above $d_l$, $\theta\ge 0$
and the low-temperature behavior of $A$ may decide between the two
most controversial spin-glass scenarios. In the droplet model
\cite{DROPLET} $\theta> 0$ and $A\sim T/L^{\theta}$ while in the RSB
scenario $\theta=0$ yielding a finite $A$ at low $T$ in the
$L\to\infty$ limit. This result corresponds to the case where there is
a finite probability density of samples with large scale excitations
of finite energy implying that $\hat{P}_v(\Delta E^*)$ for $v\sim V$
is scale independent.  The scaling relation $A\sim
T/L^{\theta}$ can be checked by numerical simulations through
measurements of $A$ (preliminary results in 2 and 3 dimensions agree
with that expression\cite{PRS2,PALASSINI}).  We remind that it holds
only for $T/L^{\theta}\ll 1$ (where the approximation (\ref{eq9}) is
valid) and does not invalidate the present analysis for $\theta<0$ if
$T$ is low enough.\\
{\it The Ising spin-glass chain.} An illustrative example of these
results is the Ising spin-glass chain
\cite{BRAY} which is described by the following Hamiltonian: \be {\cal
H}=-\sum_{i=1}^L J_i \s_i\s_{i+1}~~~~,
\label{eq18}
\ee
where the quenched disorder is distributed according to some continuous
distribution $P(J)$ with finite weight at zero coupling. The ground state
is unique (up to a global spin inversion) and has an energy $E_{\rm GS}=-\sum_i
J_i \s_i^*\s_{i+1}^*$. In the study of the low-temperature
excitations we have to distinguish between the cases with free and
periodic boundary conditions. 
In the free boundary case the minimum cost excitations correspond to
breaking the weakest bond $J^*$ and reversing all the spins on the
right or left of that bond. The energy cost of such an excitation
reads: $\Delta E^*= 2|J^*|$. Note that depending on the sample, the
length of the droplet ($v$) can range between $1$ an $L-1$. Clearly
these excitations have a finite probability of having zero cost as
$\hat P (\Delta E^*=0)\sim L\,P(J^*=0)$ which is finite by
hypothesis. Hence from our previous general arguments it follows that
at $T=0$ we obtain the universal value $G=1/3$ in agreement with the
analytical result derived in RS.  It is important to realize that
one-spin excitations cannot correspond to minimum cost excitations
when considering spins which are not at the boundaries. This is due to
the fact that, in this particular system, the ground state is not
frustrated. Exciting one interior spin would necessarily frustrate two
bonds so that the energy cost would be: $\Delta E^*=2(|J_i|+|J_k|)$ ,
which is obviously less favorable than breaking a single bond. It is
easy to check that $g_v=\frac{1}{L}$ substituted in (\ref{eq17b})
yields $A=\frac{8P(0)T(L^4-1)}{45L^3}$ in full agreement with transfer
matrix calculations shown in RS. Eqs.~(\ref{scaling},\ref{A}) with
$\gamma=d=1$  
 yield $\theta'=-1$.\\
In the periodic boundary conditions case, the situation is
different. Here we have $L$ spins and $L$ couplings, with the condition
$\s_{L+1}=\s_1$. Note that the ground state can have frustrated bonds,
so that we can distinguish between two different kinds of samples:
non-frustrated (NF) having $\prod_i J_i> 0$ and frustrated (F) having
$\prod_i J_i< 0$. In the first case the ground-state energy reads:
$E_{\rm NF}^*=-\sum_i |J_i|$ while in the second we have that the weakest
bond, let's say $J_k$, is unsatisfied so that the ground-state energy
reads: $E_{\rm F}^*=E_{\rm NF}^*+2|J_k|$. In both cases, the minimum energy
excitation will correspond to breaking the two weakest bonds, let's say
($i,k$), but the energy cost will read differently in each case: $\Delta
E_{\rm NF}^*=2(|J_i|+|J_k|), \Delta E_{\rm F}^*=2(|J_i|-|J_k|)$.  From here, it
is evident that the energy gap in the F set will be much smaller than in
the NF set. Thus, only F samples yield contributions ${\cal O}(T)$ at
low temperatures ~\cite{FOOTNOTE1}. In other words, $ \hat P(\Delta
E_{\rm F}^*=0)\neq 0$ while $ \hat P(\Delta E_{\rm NF}^*=0)= 0$ although, in
average, still $G(T=0)=1/3$ and $A\sim TL$.\\
We conclude summarizing our results. A low $T$ expansion for OPF
reveals that the only condition to get $G=1/3+{\cal O}(T)$ and $A\sim
T$ is that the ground state is unique and that the disorder average
probability distribution for the lowest excitations is gapless and
with finite weight at zero-excitation energy. This illustrates the
importance of rare samples in determining the low-temperature
properties of spin glasses.  Assuming that the statistics of the
large-scale lowest excitations determines the low $T$ thermodynamic
behavior usual scaling arguments yield $\hat{P}_{V=L^{d}}(0)\sim
L^{-\theta'}$ and $A\sim TL^{-\theta}$ providing a way to
estimate $\theta'$ and $\theta$.  In this direction, numerical
investigation of the first lowest excitations in two-dimensional spin
glasses \cite{KAWA,PRS2} would be welcome to confirm the validity of
the scenario we have put forward   through the scaling
relations (\ref{scaling}). Let us also mention that most of the
predictions presented here (in particular relations
(\ref{eq17a},\ref{eq17b},\ref{A})) can be checked with numerical
simulations for small sizes thermalized at low-temperatures
\cite{KPYHG}.  The calculations shown here open other
possibilities. Here we concentrated our attention in the study of the
parameter $q$ which gives information about the size of the
excitations. To investigate other topological aspects of the
excitations (for instance, their fractal surface exponent $d_s$) a
similar study of OPF for energy or link overlaps along the present
lines should be necessary. Further information about the topology of
the excitations can be obtained by imposing $G=1/3$ to all orders in
$T$, a result which apparently holds in spin glasses in the large
volume limit.\\
{\bf Acknowledgments}. We acknowledge enlightening discussions with
A. J. Bray and G. Parisi. We are also grateful to E. Marinari,
M. Palassini and M. Picco for useful suggestions. F.R and M.S are
supported by the M.E.C.
in Spain, project
PB97-0971 and grant AP-98 36523875 respectively.
\hspace{-2cm}

\end{multicols}
\end{document}